\documentstyle[twoside,fleqn,espcrc2,epsfig]{article}

\newcommand{\AmS}{{\protect\the\textfont2
   A\kern-.1667em\lower.5ex\hbox{M}\kern-.125emS}}

\newcommand{\be}{\begin{equation}}
\newcommand{\ee}{\end{equation}}
\newcommand{\bea}{\begin{eqnarray}}
\newcommand{\eea}{\end{eqnarray}}
\newcommand{\bdm}{\begin{displaymath}}
\newcommand{\edm}{\end{displaymath}}

\newcommand{\FF}{{\cal F}_{\pi^0\gamma^*\gamma^*}}
\newcommand{\lag}{{\cal L}}
\newcommand{\order}{{\cal O}}
\newcommand{\lapprox}{%
\mathrel{%
\setbox0=\hbox{$<$}
\raise0.6ex\copy0\kern-\wd0
\lower0.65ex\hbox{$\sim$}
}}

\title{
\vspace*{-2cm}
\hfill {\normalsize CPT-2002/P.4423} \\[-0.2cm]
\hfill {\normalsize September 2002} \\[0.5cm]
Hadronic light-by-light scattering contribution to the muon
$\lowercase{g}-2$\thanks{Talk presented at the 9th International
High-Energy Physics Conference in Quantum Chromodynamics (QCD 2002),
Montpellier, France, 2--9 July 2002.
} 
}

\author{Andreas Nyf\/feler\address{Centre de Physique Th\'{e}orique,
CNRS-Luminy, Case 907 \\  
    F-13288 Marseille Cedex~9, France \\ 
    E-mail: nyf\/feler@cpt.univ-mrs.fr
}
}

\begin{document}

\begin{abstract}
We briefly review the current status of the hadronic light-by-light
scattering correction to the muon $g-2$. Then we present our
semi-analytical evaluation of the pion-pole contribution, using a
description of the pion-photon-photon form factor based on large-$N_C$
and short-distance properties of QCD. We also sketch an effective
field theory approach to hadronic light-by-light scattering.
In view of several still unsolved problems, our conservative 
estimate for the full hadronic light-by-light scattering contribution
is $a_{\mu}^{\mbox{\tiny{LbyL;had}}} = + 8~(4) \times 10^{-10}$. 
\end{abstract}

\maketitle


\section{Introduction} 

\begin{figure}[h]
\epsfxsize=18pc 
\centerline{\epsfbox{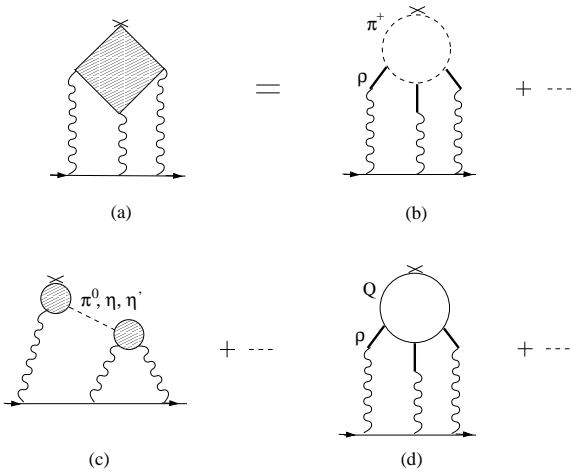}} 
\caption{The hadronic light-by-light scattering contribution to the
muon $g-2$. 
}
\label{fig:overview}
\end{figure}

The present picture of hadronic light-by-light scattering is
shown in Fig.~\ref{fig:overview} 
\vspace*{-0.3cm} 
\vspace*{-0.4cm} 
and the corresponding contributions to 
$a_\mu$ are listed in Table~\ref{tab:overview}, taking into account
the corrections made in the two full
evaluations~\cite{HKS_corr,BPP_corr}, after we had discovered the sign
error in the pion-pole contribution~\cite{KN_pion,a_mu_EFT}.
\begin{table}[hbt]
\caption{Contributions to $a_{{\scriptscriptstyle
\mu}} (\times 10^{10})$ according to Fig.~\ref{fig:overview}. The last
column gives the result when no form factors are used in the couplings
to the photons.} 
\label{tab:overview}   
\begin{center}
\renewcommand{\arraystretch}{1.1}
\begin{tabular}{|l|r@{.}l|r@{.}l|c|c|}
\hline
Type &
\multicolumn{2}{|c|}{{Ref.~\cite{HKS_corr}}} & 
\multicolumn{2}{|c|}{{Ref.~\cite{BPP_corr}}} & 
Ref.~\cite{KN_pion} & \\ 
\hline  
(b)                & -0 & 5(0.8) & -1 & 9(1.3) & & -4.5 \\ 
(c) & 8 & 3(0.6) & 8 & 5(1.3) & 8.3(1.2)    & $+\infty$ \\
$f_0, a_1$               & 0 & 174$^{\rm a}$ & -0 & 4(0.3) &   
&  \\
(d) & 1 & 0(1.1) & 2 & 1(0.3) &    & $\sim 6$ \\
\hline
Total & 9 & 0(1.5) & 8 & 3(3.2) & 8(4)$^{\rm b}$ &  \\ 
\hline
\end{tabular}
\end{center}
$^{\rm a}$~Only $a_1$ exchange. \\
$^{\rm b}$~Our estimate, using
Refs.~\cite{HKS_corr,BPP_corr,KN_pion}. 
\vspace*{-0.4cm}
\end{table}
There are three classes of contributions to the hadronic four-point
function [Fig.~\ref{fig:overview}(a)], which can be understood from an
effective field theory (EFT) analysis of hadronic light-by-light
scattering~\cite{EdeR_94,a_mu_EFT}: (1) a charged pion loop
[Fig.~\ref{fig:overview}(b)], where the coupling to photons is dressed
by some form factor ($\rho$-meson exchange, e.g.\ via vector meson
dominance (VMD)), (2) the pseudoscalar pole diagrams
[Fig.~\ref{fig:overview}(c)] together with the exchange of heavier
resonances ($f_0, a_1, \ldots$) and, finally, (3) the irreducible part
of the four-point function which was modeled in
Refs.~\cite{HKS_corr,BPP_corr} by a constituent quark loop dressed
again with VMD form factors [Fig.~\ref{fig:overview}(d)]. The latter
can be viewed as a local contribution $\bar\psi
\sigma^{\mu\nu} \psi F_{\mu\nu}$ to $g-2$. The two 
groups~\cite{HKS_corr,BPP_corr} used similar, but not identical models
which explains the slightly different results for the dressed charged
pion and the dressed constituent quark loop, although their sum seems
to cancel to a large extent and the final result is essentially given
by the pseudoscalar exchange diagrams. Since the models used in
Refs.~\cite{HKS_corr,BPP_corr} are not fully consistent with QCD, we
take the difference of the results as indication of the error coming
from the model dependence.

On the other hand, we will show in Section~\ref{sec:pionpole} that the
pseudoscalar contribution now seems under control, due to our
semi-analytical calculation~\cite{KN_pion}, based on a
pion-photon-photon form factor $\FF$ which fulfills the
relevant QCD short-distance constraints~\cite{KN_VAP}, in contrast to
the form factors used in Refs.~\cite{HKS_corr,BPP_corr}. These
findings are also corroborated by an EFT and large-$N_C$
analysis of $a_{\mu}^{\mbox{\tiny{LbyL;had}}}$~\cite{a_mu_EFT} which
allows to calculate the leading and next-to-leading logarithms 
(Sec.~\ref{sec:EFT}).


\section{Pion-pole contribution}
\label{sec:pionpole} 

The contribution from the neutral pion intermediate state is given by
a two-loop integral that involves the convolution of two
pion-photon-photon transition form factors, see
Fig.~\ref{fig:overview}(c). Since no data on the doubly off-shell form
factor $\FF(q_1^2,q_2^2)$ is available, one has to resort to
models. In order to proceed with the analytical evaluation of the
two-loop integrals, we considered a certain class of form factors
which includes the ones based on large-$N_C$ QCD that we studied in
Ref.~\cite{KN_VAP}.  For comparison, we have also used a vector meson
dominance (VMD) and a constant form factor, derived from the
Wess-Zumino-Witten (WZW) term.  For all these form factors we could
perform {\it all} angular integrations in the two-loop integrals
analytically~\cite{KN_pion}.

In large-$N_C$ QCD, the pion-photon-photon form factor is described by
a sum over an infinite set of narrow vector resonances, involving
arbitrary couplings, although there are constraints at long and short
distances. The normalization is given by the WZW term, $\FF(0,0) = -
N_C / (12 \pi^2 F_\pi)$, whereas the OPE tells us that
\bea
&& \lim_{\lambda\to \infty}\,\FF(\lambda^2 q^2, (p-\lambda
q)^2)  \nonumber \\  
&& = \frac{2}{3}\,\frac{F_\pi}{q^2}\,\bigg\{ \frac{1}{\lambda^2}\, 
+ \frac{1}{\lambda^3} \frac{q \cdot p}{q^2} 
+\,{\cal O}\left({1 \over \lambda^4}\right)\bigg\}\,. 
\label{OPE_FF}
\eea
In the following, we consider the form factors that are obtained by
truncation of the infinite sum in large-$N_C$ QCD to one (lowest meson
dominance, LMD), and two (LMD+V), vector resonances per
channel, respectively:
\be
\FF^{\mbox{{\tiny LMD}}}(q_1^2,q_2^2) =   \!{F_\pi \over 3} {
q_1^2 + q_2^2 - c_V  \over (q_1^2 - M_V^2) (q_2^2 - M_V^2) }, 
\label{FF_LMD}
\ee
\bea
&&\hspace*{-0.7cm}\FF^{\mbox{{\tiny LMD+V}}}(q_1^2,q_2^2)\!=\!{F_\pi
\over 3} \Bigg\{\!\bigg(\!\!q_1^2\!q_2^2
(q_1^2\!+\!q_2^2)\!+\!h_1\!(q_1^2\!+\!q_2^2)^2 
\nonumber \\
&&\!\!+ h_2q_1^2 q_2^2 + h_5 (q_1^2\!+\!q_2^2) + 
h_7 \bigg) \bigg/ \bigg( (q_1^2 - M_{V_1}^2) \nonumber \\
&& \times (q_1^2 - M_{V_2}^2) (q_2^2 - M_{V_1}^2) (q_2^2 -
M_{V_2}^2) \bigg) \Bigg\}, \label{FF_LMD+V} 
\eea
with the constants $c_V \,= \, N_C M_V^4 / (4\pi^2 F_\pi^2)$ and $h_7
\,= \, - N_C M_{V_1}^4 M_{V_2}^4 / (4\pi^2 F_\pi^2)$.  The parameters
$h_1,h_2,$ and $h_5$ in the LMD+V form factor are not fixed by the
normalization and the leading term in the OPE. We have determined
these coefficients phenomenologically~\cite{KN_VAP,KN_pion}.  In
particular, $\FF(\!-Q^2,\!0)$ with one photon on-shell behaves like
$1/Q^2$ for large spacelike momenta, $Q^2\!=\!-q^2$. Whereas the LMD
form factor does not have such a behavior, it can be reproduced with
the LMD+V ansatz, provided that $h_1\!=\!0$. A fit to the data yields
moreover $h_5\!=\!6.93 \pm 0.26~\mbox{GeV}^4$. Analyzing the
experimental data for the decay $\pi^0\!\to\!e^+ e^-$ leads to the
loose bound $|h_2|\!\lapprox\!20~\mbox{GeV}^2$.

Note that the usual VMD form factor $\FF^{\mbox{\tiny
VMD}}(q_1^2,q_2^2) \sim 1/ [(q_1^2 - M_V^2) (q_2^2 - M_V^2)]$ does
{\it not} correctly reproduce the OPE in Eq.~(\ref{OPE_FF}).

After performing the angular integrations, the pion-exchange
contribution to $a_\mu$ can be written as a two-dimensional integral
representation, where the integration runs over the moduli of the
Euclidean momenta
\bea
a_{\mu}^{\mbox{\tiny{LbyL;$\pi^0$}}} & = & \int_0^\infty d Q_1 
\int_0^\infty d Q_2 \nonumber \\
&& \quad \times \sum_i w_i(Q_1,Q_2) \ f_i(Q_1,Q_2) ,  
\eea
with universal [for the above class of form factors] weight functions
$w_i$ (rational functions, square roots and
logarithms)~\cite{KN_pion}. The dependence on the form factors resides
in $f_i$. In this way we could separate the generic features of the
pion-pole contribution from the model dependence. This is not possible
anymore in the final analytical result (as a series expansion) for
$a_{\mu}^{\mbox{\tiny{LbyL;$\pi^0$}}}$  in Ref.~\cite{Blokland_etal}. 
One has to keep in mind that there is an intrinsic uncertainty in the
form factor of $10 - 30~\%$, furthermore the VMD form factor used in
that reference has the wrong high-energy behavior.

The weight functions $w_i$ in the main contribution are positive and
peaked around momenta of the order of $0.5~\mbox{GeV}$. There is,
however, a tail in one of these functions, which produces for the
constant WZW form factor a divergence of the form $\ln^2\!\Lambda$ for
some UV-cutoff $\Lambda$. Other weight functions have positive and
negative contributions in the low-energy region, which lead to a
strong cancellation in the corresponding integrals.

In Table~\ref{tab:api_models} we present the numerical results for the
different form factors.  All form factors lead to very similar results
(apart from WZW). Judging from the shape of the weight functions
described above, it seems more important to correctly reproduce the
slope of the form factor at the origin and the available data at
intermediate energies. On the other hand, the asymptotic behavior at
large $Q_i$ seems not very relevant.  The results for the LMD+V form
factor are rather stable under the variation of the parameters, except
for $h_2$. If all other parameters are kept fixed, our result changes
in the range $|h_2| < 20~\mbox{GeV}^2$ by $\pm 0.9 \times 10^{-10}$
from the value for $h_2 = 0$. \\
\begin{table}[b]
\caption{Results for 
$a_{{\scriptscriptstyle \mu}}^{\mbox{\tiny{LbyL;$\pi^0$}}}$ for the
different form factors. In the WZW model we used a cutoff of
$1~\mbox{GeV}$ in the divergent contribution. 
}
\label{tab:api_models}   
\begin{center}
\renewcommand{\arraystretch}{1.1}
\begin{tabular}{|l|r@{.}l|}
\hline
Form factor &
\multicolumn{2}{|c|}{{$a_{{\scriptscriptstyle
\mu}}^{\mbox{\tiny{LbyL;$\pi^0$}}} \times 10^{10}$}}    
\\ 
\hline  
WZW     & \hspace*{0.5cm} 12 & 2 \\  
VMD     &  5 & 6 \\ 
LMD     & 7 & 3 \\
LMD+V ($h_2 = 0~\mbox{GeV}^2$)
        & 5 & 8 \\
\hline
\end{tabular}
\end{center}
\end{table}
\hspace*{2mm}Thus, with the LMD+V form factor, we get 
\be 
a_{\mu}^{\mbox{\tiny{LbyL;$\pi^0$}}} = + 5.8~(1.0) \times 10^{-10}
\, , 
\ee
where the error includes the variation of the parameters and the
intrinsic model dependence. A similar short-distance analysis in the
framework of large-$N_C$ QCD and including quark mass corrections for
the form factors for the $\eta$ and $\eta^\prime$ was beyond the scope
of Ref.~\cite{KN_pion}. We therefore used VMD form factors fitted to
the available data for $\FF(-Q^2,0)$ to obtain our final estimate 
\bea
a_{\mu}^{\mbox{\tiny{LbyL;PS}}} & \equiv & 
a_{\mu}^{\mbox{\tiny{LbyL;$\pi^0$}}}
+ a_{\mu}^{\mbox{\tiny{LbyL;$\eta$}}}\vert_{\mbox{\tiny VMD}} +
a_{\mu}^{\mbox{\tiny{LbyL;$\eta^\prime$}}}\vert_{\mbox{\tiny VMD}}
\nonumber \\
& = & + 8.3~(1.2) \times 10^{-10} \, . 
\eea
An error of 15~\% for the pseudoscalar pole contribution seems
reasonable, since we impose many theoretical constraints from long 
and short distances on the form factors. Furthermore, we use
experimental information whenever available. 


\section{EFT approach to $a_{\mu}^{\mbox{\tiny{LbyL;had}}}$}
\label{sec:EFT}

In Ref.~\cite{a_mu_EFT} we discussed an EFT approach to hadronic
light-by-light scattering based on an effective Lagrangian that
describes the physics of the Standard Model well below 1~GeV. It
includes photons, light leptons, and the pseudoscalar mesons and obeys
chiral symmetry and $U(1)$ gauge invariance.

The leading contribution to $a_{\mu}^{\mbox{\tiny{LbyL;had}}}$, of
order $\order(p^6)$, is given by a (finite) loop of charged pions with
point-like electromagnetic vertices.  Since this contribution involves
a loop of hadrons, it is subleading in the large-$N_C$ expansion.

\begin{figure}[h]
\epsfxsize=18pc 
\centerline{\epsfbox{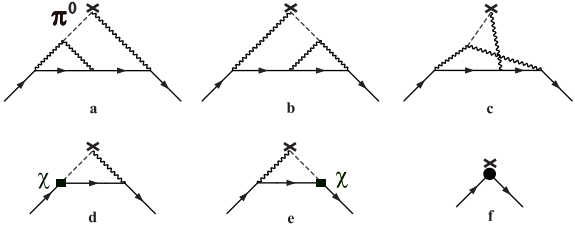}} 
\caption{The graphs contributing to
$a_{\mu}^{\mbox{\tiny{LbyL;$\pi^0$}}}$ at lowest order in the effective
field theory. 
\label{fig:EFT}}
\end{figure}

At order $p^8$ and at leading order in $N_C$, we encounter the
divergent pion-pole contribution, diagrams (a) and (b) of
Fig.~\ref{fig:EFT}, involving two WZW vertices.  
The diagram (c) is actually finite. The divergences of the triangular
subgraphs in the diagrams (a) and (b) are removed by inserting the
counterterm $\chi$ from the Lagrangian $\lag^{(6)} = (\alpha^2 / 4
\pi^2 F_0) \ \chi \ {\overline\psi} \gamma_\mu \gamma_5 \psi \,
\partial^\mu \pi^0 + \cdots$, see the one-loop diagrams (d) and
(e). Finally, there is an overall divergence of the two-loop diagrams
(a) and (b) that is removed by a local counterterm, diagram (f).
Since the EFT involves such a local contribution, we will not be able
to give a precise numerical prediction for
$a_{\mu}^{\mbox{\tiny{LbyL;had}}}$.

Nevertheless, it is interesting to consider the leading and
next-to-leading logarithms that are in addition enhanced by a factor
$N_C$ and which can be calculated using the renormalization
group~\cite{a_mu_EFT}. The EFT and large-$N_C$ analysis tells us that 
\bea
&&\!\!\!\!\!\!\!\!a_{\mu}^{\mbox{\tiny{LbyL;had}}} = 
\left( {\alpha \over \pi} \right)^3  \Bigg\{
f\left({m_{\pi^\pm} \over m_\mu}, {m_{K^\pm} \over m_\mu}\right)
 \nonumber \\
&&\!\!\!\!\!\!\!\!\!+ N_C \left( {m_\mu^2 \over 16 \pi^2
F_\pi^2} {N_C \over 3} \right)
\left[ \ln^2 {\mu_0 \over m_\mu} + c_1 \ln {\mu_0 \over m_\mu} + c_0
\right]  \nonumber \\ 
&&\!\!\!\!\!\!\!\!\!+ \order \left(\!{m_\mu^2 \over \mu_0^2} \times
\mbox{log's}\!\right) + \order \left(\!{m_\mu^4 \over \mu_0^4} 
N_C \times \mbox{log's}\!\right)\!\!\Bigg\},  \label{a_mu_EFT_N_C}
\eea
where $f(m_{\pi^\pm}\!/\!m_\mu,\!m_{K^\pm}\!/\!m_\mu)\!=\!\!-0.038$
represents the charged pion and kaon-loop that is formally of order
one in the chiral and $N_C$ counting  and $\mu_0$ denotes some
hadronic scale,  
e.g.\ $M_\rho$.  The coefficient ${\cal C}$ of the log-square term in
the second line is universal and of order $N_C$, since
$F_\pi\!=\!\order(\sqrt{N_C})$.  

Unfortunately, although the logarithm is sizeable, in
$a_{\mu}^{\mbox{\tiny{LbyL;$\pi^0$}}}$ there occurs a cancellation
between the log-square and the log-term. If we fit our result for the
VMD form factor for large $M_\rho$ to an expression as given in
Eq.~(\ref{a_mu_EFT_N_C}), we obtain
\bea
a_{\scriptscriptstyle{\mu;\mbox{\tiny{VMD}}}}^{\mbox{\tiny{LbyL}}; \pi^0}
& \doteq &\!\!\left( {\alpha \over \pi} \right)^3 {\cal C} 
\ \ \left[ \ln^2 {M_\rho \over m_\mu} + c_1 \ln {M_\rho \over m_\mu} + c_0
\right]  \nonumber \\
& \stackrel{\mbox{\tiny{Fit}}}{=} &\!\!\left( {\alpha \over \pi}
\right)^3 {\cal C}  
\ \ \left[ 3.94 - 3.30 + 1.08 \right] \nonumber \\
& = & \hspace*{0cm}\!\!\left[ 12.3 - 10.3 + 3.4 \right] \times 10^{-10}
\nonumber \\
& = &\!\!5.4 \times 10^{-10} \, , 
\eea
which is confirmed by the analytical result of
Ref.~\cite{Blokland_etal} (setting for simplicity $m_{\pi^0} =
m_\mu$): $a_{\scriptscriptstyle{\mu;
\mbox{\tiny{VMD}}}}^{\mbox{\tiny{LbyL}}; \pi^0} = [12 - 8.0 + 1.7]
\times 10^{-10} = 5.7 \times 10^{-10}$. This 
cancellation is now also visible in the revised version of
Ref.~\cite{Ramsey-Musolf_Wise}. In that paper the remaining parts
of $c_1$ have been calculated: $c_1 = - 2 \chi(\mu_0) / 3 + 0.237 =
-0.93^{+0.67}_{-0.83}$, with our conventions for $\chi$ and
$\chi(M_\rho)_{{\rm exp}} = 1.75^{+1.25}_{-1.00}$.   


\section{Conclusions}

The pseudoscalar pole con\-tri\-bu\-tion
$a_{\mu}^{\mbox{\tiny{LbyL;PS}}}$ seems to be under control at the
15~\% level. Moreover, the EFT and large-$N_C$ analysis shows a
systematic approach to $a_{\mu}^{\mbox{\tiny{LbyL;had}}}$ and yields
the leading and next-to-leading logarithmic terms, enhanced by a
factor $N_C$, although these terms tend to cancel each other. However,
there remains the issue of the other contributions, i.e.\ the dressed
charged pion and the dressed constituent quark loop, see
Fig.~\ref{fig:overview} and Table~\ref{tab:overview}, where model
calculations lead to slightly different results.  Taking these
uncertainties into account by adding the errors linearly, my estimate
for the full hadronic light-by-light scattering contribution reads:
\be
a_{\mu}^{\mbox{\tiny{LbyL;had}}} = +\,8~(4) \times 10^{-10} \, . 
\ee


\section*{Acknowledgments}

I would like to thank the organizers of QCD~02, in particular S.\
Narison, for providing such a pleasant atmosphere and the
Schweizerischer Nationalfonds for financial support.


\end{document}